Nature Review

# The role of black holes in galaxy formation and evolution

A. Cattaneo[1,2], S. M. Faber[3], J. Binney[4], A. Dekel[5], J. Kormendy[6], R. Mushotzky[7], A. Babul[8], P. N. Best[9], M. Brüggen[10], A. C. Fabian[11], C. S. Frenk[12], A. Khalatyan[13], H. Netzer[14], A. Mahdavi[15], J. Silk[4], M. Steinmetz[1] & L. Wisotzki[1]

**[1]Astrophysikalisches Institut Potsdam, An der Sternwarte 16, 14482 Potsdam, Germany**

**[2]Observatoire de Lyon, Université de Lyon 1, 9 avenue Charles André, 69561 Saint Genis Laval cedex, France**

**[3]University of California Observatories/Lick Observatory, University of California, Santa Cruz, California 95064, USA**

**[4]Department of Physics, University of Oxford, Keble Road, Oxford OX1 3RH, UK**

**[5]Racah Institute of Physics, The Hebrew University, Jerusalem 91904, Israel**

**[6]Department of Astronomy, University of Texas, Austin, Texas 78712, USA**

**[7]Goddard Space Flight Center, NASA, Greenbelt, Maryland 20771, USA**

**[8]Department of Physics and Astronomy, University of Victoria, Elliot Building, 3800 Finnerty Road, Victoria, British Columbia V8P 1A1, Canada**

**[9]Institute for Astronomy, Royal Observatory, Blackford Hill, Edinburgh EH9 3HJ, UK**

**[10]Jacobs University Bremen, Campus Ring 1, 28759 Bremen, Germany**

**[11]Institute of Astronomy, University of Cambridge, Madingley Road, Cambridge CB3 0HA, UK**

**[12]Institute for Computational Cosmology, University of Durham, South Road, Durham DH1 3LE, UK**

**[13]Observatoire Astronomique Marseille-Provence, 38 rue Frédéric Joliot-Curie, 13388 Marseille cedex 13, France**

**[14]Wise Observatory, University of Tel Aviv, 69978 Tel Aviv, Israel,**

**[15]Department of Physics and Astronomy, San Francisco State University, 1600 Holloway Avenue, San Francisco, California 94132, USA**

**Virtually all massive galaxies, including our own, host central black holes ranging in mass from millions to billions of solar masses. The growth of these black holes releases vast amounts of energy that powers quasars and other weaker active galactic nuclei. A tiny fraction of this energy, if absorbed by the host galaxy, could halt star formation by heating and ejecting ambient gas. A central question in galaxy evolution is the degree to which this process has caused the decline of star formation in large elliptical galaxies, which typically have little cold gas and few young stars, unlike spiral galaxies.**

Galaxies come in two basic types: ‘football-shaped’ ellipticals and ‘disk-shaped’ spirals (Fig. 1). Spirals contain plenty of cold gas, which forms stars, whereas the gas in ellipticals is too hot to form stars. Thus, ellipticals lack the young blue stars that are usually seen in spirals, and are generally quite red. Spirals also have central bulges structurally resembling miniature ellipticals. Owing to this similarity, we use the term ‘bulges’ for bulges within spirals and for ellipticals indiscriminately.

Each bulge contains a central black hole, whose mass is proportional to the bulge stellar mass[1–5], $M_{BH} \approx 0.001 M_{bulge}$. Black holes and bulges also formed at about the same epoch in the lifetime of the Universe[6,7]. These observations imply that the formation of black holes and the formation of bulges are closely linked. Matter falling onto a black hole releases a huge amount of energy[8], of the order of 10% of the rest mass energy, $E = mc^2$, mainly in the form of photons but also in the form of radio-luminous jets of charged particles[9,10]. Even a tiny fraction (<1%) of the energy released within each bulge could heat and blow away its entire gas content, thus explaining the lack of star formation in bulges.

The theorist’s goal is to understand these observations in a cosmological context. In the standard picture[11–13], most of the Universe is composed of dark matter, whose nature is unknown. Protons, electrons and neutrons, which compose gas and stars, make up the rest. They interact with dark matter purely through gravity, which determines the evolution of the Universe on large scales. The Universe emerged from the Big Bang with small inhomogeneities. These eventually grew into lumps, called haloes, by attracting surrounding matter gravitationally (Fig. 2). The competition between radiative cooling and gravitational heating determines the fate of gas in these haloes[14–16]. In low-mass haloes, cooling dominates. Galaxies grow through the accretion of gas that falls to the centre in cold flows[17,18], settles into disks[19] (but see refs 20, 21), and forms stars. However, when the halo mass grows above a critical value of about $10^{12}$ solar masses[18], heating dominates, and the gas no longer accretes onto galaxies. Halo mergers form large haloes that contain tens or even hundreds of galaxies, called groups or clusters, respectively. Galaxy mergers within haloes transform disks into bulges[22] and are the only opportunity for galaxies to grow after they have ceased to accrete gas.

In mergers of galaxies that are still accreting gas, the gas falls to the centre, triggers starbursts, and is often observed to feed the rapid growth of black holes[23,24], although, observationally, the connection between active galactic nuclei (AGN) and mergers remains controversial[25–27]. Black holes respond to this fuelling by feeding energy back to the surrounding gas. This energy produces winds[28–33], which may accelerate the star formation rate by compressing the gas[34]. In the most dramatic scenario, all the gas is blown away, so black hole growth and star formation suddenly terminate[35–39]. Computer simulations suggest that this ‘quenching’ is necessary to explain why ellipticals are red[40,41]. The chemical composition of giant ellipticals implies that the star formation epoch was brief. This is another reason for considering a mechanism that could 'quench' star formation abruptly[42].

In galaxies that have ceased to accrete gas, the main epoch of star formation and black hole accretion is over, but even weak AGN can open large cavities in the hot gas through the mechanical action of their jets[43–45]. In this 'maintenance mode', black holes couple to the hot gas and adjust their residual accretion rates to provide the energy that is needed to maintain it at constant temperature[46–48], either through a continuous series of minor events[44,45,49], or through episodic quasar activity[50,51]. The strongest evidence for this loop cycle is in galaxy clusters. Its action reduces considerably the rate at which gas cools and accretes onto the central galaxies[52,53].

These discoveries have lead to a richer picture of galaxy formation, in which black holes play a major part. Modelling these effects is essential to understand the masses, colours[54–56] and structural properties[57] of ellipticals within a cosmological theory of galaxy formation.

## Black holes in the formation of red ellipticals

Black hole masses are tightly related to the stellar mass[3] and the stellar velocity dispersion[4,5] within the host bulges. These correlations imply a causal link between the formation of black holes and the formation of bulges, but they can be interpreted in two ways.

In the first interpretation, black hole accretion and star formation occur together because they both feed from the same gas, brought to the centre by gas-rich mergers and disk instabilities. Black hole accretion terminates when star formation has used up all the gas. The correlation between black hole growth and starbursts observed in ultraluminous infrared galaxies[24], quasars[58] and nearby Seyfert galaxies[59] supports this scenario, explaining why the black hole mass is proportional to the bulge mass.

In the second interpretation, star formation terminates when the black hole blows all the gas outside its host galaxy[35–38]. Feedback requires a minimum power and thus a minimum mass because, for a given black hole mass, there is a maximum AGN luminosity, called the Eddington limit, above which the radiation-pressure force outwards exceeds the gravitational force inwards, suppressing the gas flow onto the black hole. The velocity dispersion is the bulge property that is most closely linked to the black hole because it determines the depth of the potential well from which the gas has to be expelled, and thus the minimum black hole mass for feedback.

The efficiency with which AGN can transfer energy to the surrounding gas determines which picture is closer to reality. Photons and jets from AGN can drive winds in two ways. They can heat the gas and cause it to expand (thermal ‘energy-driven’ winds) or they can push it out (pressure ‘momentum-driven’ winds). Photons heat the gas by photoionizing metals such as iron, which retain their inner electrons even at high temperature, and by Compton scattering. Radiation pressure on ionized gas is only important in the immediate proximity of black holes, where it determines the Eddington limit, but radiation pressure on dust is important even on galactic scales because dust has a high absorption cross-section. Absorption by resonance lines is another mechanism that transfers momentum to the gas, and may explain the high-speed winds in broad absorption line quasars[33]. Jets can produce ‘energy-driven’ winds via shock heating and ‘momentum-driven’ winds via ram pressure.

All these processes contain large inefficiencies, which are difficult to quantify: metals that retain some electrons even at high temperature are a small fraction of the atoms in the gas; the photoionization cross-section is large only in a narrow frequency band; Compton scattering transfers only a small fraction of the photon energy to an electron; and jets tend to escape from their host galaxies and to deposit most of their energy outside. The transfer of momentum from the photons to the gas by way of radiation pressure is the only process that can be almost 100% efficient, but dust must cover a large solid angle. The low momentum of photons with respect to their energy also limits the mass that can be ejected through this mechanism, as photons are massless particles. If the momentum in the photons radiated by an AGN was transferred to the gas with 100% efficiency, this momentum could eject a gas mass equal to at most ~10% of the bulge stellar mass, which is the mass of the gas that is typically left over at the end of a gas-rich merger. Thermal-wind[35,39] and radiation-pressure-driven-wind[36–38] models have been used to compute the relation between black hole mass and bulge velocity dispersion. However, this is not a discriminating test because both cases can match the data.

Optical/ultraviolet[28,29] and X-ray[30,31] spectroscopy confirm that quasars can accelerate winds with speeds of thousands to tens of thousands of kilometres per second. Their variability on short timescales suggests that these are nuclear rather than galactic winds, but some kiloparsec-scale winds are observed[29]. Integral-field spectroscopy has also detected bipolar winds with speeds of thousands of kilometres per second aligned with the jets of high-redshift radio galaxies[32].

The mere existence of AGN winds is no proof that they ‘quench’ star formation, but observations of post-starburst galaxies find that two-thirds of them contain winds with speeds of 500–2,000 km $s^{-1}$ (ref. 60). These speeds are higher than the wind speeds usually found in starbursts. They, therefore, suggest a quasar origin and a probable role of quasars in the ‘quenching’ of post-starburst galaxies.

In the Sloan Digital Sky Survey (SDSS), which probes nearby galaxies, star-forming, 'active' and 'passive' ellipticals delineate a sequence from blue to red on the galaxy colour–mass diagram[61]. The interpretation is that when the growth of the black hole is activated, the star formation rate declines. However, in the overall population of SDSS galaxies, the star formation decline appears to be gradual[62] and is not linked to any dramatic event. The situation may be different at high redshift, where starbursts and quasars were more common and more powerful, and where star formation in the progenitors of giant ellipticals lasted for less than a gigayear (ref. 42).

## Black holes in galaxy clusters

The gas in massive galaxies, groups and clusters is hot and radiates copiously in X-rays. The problem of explaining why this gas does not quickly cool off is known as the 'cooling flow' problem. This problem has been heavily investigated in galaxy clusters, where the observational constraints are particularly strong[52,53].

X-ray groups and clusters fall into two categories: systems in which the X-ray surface brightness increases steeply towards the centre—that is, 90% of the X-ray-selected groups and clusters with halo mass ($M_{\rm halo}$) $\leq 10^{14}$ solar masses, and 50% of the clusters with $M_{\rm halo} \geq 10^{14}$ solar masses[63]—and systems with shallower surface brightness profiles. The gas luminosity per unit volume is equal to $n^2\Lambda$, where $n$ is the gas density, and $\Lambda$ is the cooling function; $\Lambda$ depends on the gas temperature $T$ and on the gas chemical composition. X-ray spectroscopy shows that $T$ is always of the order of the virial temperature $T_{\rm vir}$ at which the gas is in equilibrium with gravity. Therefore, the difference between the two types of cluster must be in their density profiles.

In the first type of cluster, the gas has a high central density and radiates its thermal energy on a timescale $t_{\rm cool} \approx (3/2)kT/n^2\Lambda$, where $k$ is the Boltzmann constant, that is usually less than a gigayear over much of the cluster core. These clusters are called 'cool-core' clusters because $T$ decreases towards the centre. However, cool-core clusters are not 'cold core' clusters: the temperature drops towards the centre by only a factor of three. From the absence or weakness of the soft X-ray line Fe XVII, one infers that the amount of gas that cools radiatively below this temperature is ten times less than expected from how much heat is lost to X-rays[52,53]. As the gas radiates but does not cool, there must be a compensating energy-injection mechanism.

Further evidence comes from the relation between X-ray luminosity ($L_{\rm X}$) and X-ray temperature ($T$). At $T \geq 3$ keV, where bremsstrahlung is the main radiation mechanism, the X-ray luminosity is $L_{\rm X} \propto n^2 T_{\rm vir}^{1/2} r_{\rm halo}^3$, and $T_{\rm vir} \propto M_{\rm halo}/r_{\rm halo}$, where $r_{\rm halo}$ is halo radius. If $n$ scaled with the halo density, which is proportional to the mean density of the Universe, then all clusters should have the same $n$ and this equation predicts $L_X \propto T_{vir}^2$. Spectroscopy confirms that $T \approx T_{vir}$, but the data find a different relation $L_X \propto T_{vir}^3$, which becomes even steeper at $T_{vir} \leq 3$ keV (ref. 64) because $n$ decreases at low masses, although with considerable scatter[65–67]. Having lower density for a given temperature implies having higher entropy, measured by $K = kT/n^{2/3}$, and the only way to increase the entropy is through heating.

Clusters have typical entropy excesses of $\Delta K \approx 100$ keV cm$^2$ at $0.1r_{halo}$ (ref. 68; Fig. 3). These excesses weigh more heavily on smaller clusters, which have lower absolute entropies, but higher entropies relative to theoretical expectations. This problem is common to both cool-core and non-cool-core clusters, and affects a large fraction of the intracluster medium. The quasar winds invoked to quench star formation in the progenitors of giant ellipticals could solve this entropy problem by preheating the intergalactic gas destined to become the intracluster medium[69–71], but they cannot solve the cooling-flow problem in the central regions of cool-core clusters. In these systems, which have $K < 100$ keV cm$^2$, the cooling time is so short (<0.1 Gyr; Fig. 3) that one needs heating at least every 0.1 Gyr today to maintain these systems in their current state. This need for regular heating clashes with the scarcity of quasars in the low-redshift Universe.

However, weaker AGN—that is, 'edge-darkened' radio galaxies—show up in 70% of the central dominant (cD) galaxies of cool-core clusters[72,73]. Their activity pattern differs from the erratic behaviour of quasars and is closer to a constant string of minor outbursts. In many low-accretion-rate AGN, the gas surrounding the black hole is not dense enough to radiate efficiently, and nearly all the released energy may instead be channelled into jets[74]. The very poor optical luminosity of these objects is partly the reason why their importance had long been underestimated.

The importance of radio galaxies began to emerge after cavities were discovered in the X-ray gas of the Perseus cluster[43,45] (Fig. 4) and other clusters with substantially weaker AGN—for example, the Virgo cluster[44]. The cavities are regions where the jet radio-synchrotron-emitting plasma has displaced the ambient X-ray-emitting plasma. Such cavities are present in ≥70% of cool-core clusters[75]. They are usually regions of enhanced synchrotron emission, although some lack high-frequency radio emission ('ghost' cavities), presumably because they are old and depleted of electrons with energies $>10^5 m_e c^2$, where $m_e$ is the electron mass.

From the volume of the cavities ($V_{cav}$) and the pressure of the intracluster medium ($p_{ICM}$) it is possible to estimate the work that the jets had to do to create them. This work equals $p_{ICM}V_{cav}$ for 'quasi-static' (that is, highly subsonic) inflation. However, as cavities are Rayleigh-Taylor unstable, their formation timescale cannot be much longer than the sound crossing time or they would break apart before they are formed. It cannot be much shorter than the sound crossing time either. If it were, cavities would be surrounded by strong shocks, observed only in very few objects. As cavities are not inflated quasi-statically, the work that the expanding radio lobes do on the ambient gas must exceed $p_{\rm ICM}V_{\rm cav}$. Part of this extra work excites shocks, waves and other disturbances, all of which could heat the intracluster medium. Analytic calculations[76] and numerical simulations[77] indicate that this work could be up to $10p_{\rm ICM}V_{\rm cav}$.

In addition to the work that radio lobes do on the ambient gas, there is also the energy of the relativistic particles inside the radio lobes themselves ('cosmic rays'). The internal energy of the radio-emitting plasma is $1/(\gamma-1)p_{\rm cav}V_{\rm cav}$, where cavity pressure $p_{\rm cav} \geq p_{\rm ICM}$ and $\gamma$ is the plasma adiabatic index ($\gamma = 4/3$ is the value that is normally assumed for a relativistic plasma). This energy could heat the intracluster medium, too, if the synchrotron emitting plasma and the X-ray emitting gas eventually mixed, although observations show that cavities tend to survive in the intracluster medium for a very long time (see, for example, ref. 76).

The minimum energy needed to produce the observed cavities ($E_{\rm cav}$) is obtained by adding the work done on the ambient gas for quasi-static inflation and the cosmic ray energy. This sum gives $E_{\rm cav} \geq 4p_{\rm ICM}V_{\rm cav}$. Cavity observations find that this energy is equal to the energy radiated in X-rays in a sound crossing time, to within a factor of four (ref. 46; Fig. 5a). Thus, the energy that jets put into cavities is about equal to the energy needed to offset cooling. This near-equality, which extends over four orders of magnitude, suggests a self-regulation mechanism. This is possible, because a black hole acts as a thermostat that senses the entropy of the gas at the boundary of its gravitational sphere of influence, determined by the Bondi radius[78] $r_{\rm Bondi} \approx GM_{\rm BH}/c_{\rm s}^2 \approx 10$–$100$ pc, where the gas infall speed equals the sound speed $c_{\rm s}$ (here $G$ is the gravitational constant). For spherical accretion[78], $M_{\rm BH}$ and the entropy at $r_{\rm Bondi}$ entirely determine the black hole accretion rate. If the power injected into the cavities is proportional to the black hole accretion rate computed from a spherical model—an assumption directly verified by observational data[79] (Fig. 5b)—then, the more the gas cools down and the central entropy decreases, the more the jet power and the heating rate go up.

The importance of jet heating is also shown by the entropy of the intracluster medium, which is higher in clusters with extended radio sources than in clusters with point-like radio sources, where jets have not yet propagated and heated the gas[80]. Properties of clusters with point-like sources are in fact consistent with the non-heating prediction, $L_{\rm X} \propto T_{\rm vir}^2$.

The problem is how the energy injected into clusters is converted into heat. Strong shocks seem the most natural mechanism, but in fact the X-ray-bright rims of the radio lobes in the Perseus cluster[45] (Fig. 4) and other radio galaxies are cooler than their surroundings. Moreover, if the heated region were substantially smaller than the cluster core, convection would set in. Instead, the entropy profiles of clusters are shallow but not flat (Fig. 3). This indicates that the central gas is stable against convection, even though the gas metallicity profiles are broader than the starlight of the cD galaxies[53]. Different stages in the life of a radio source could explain the scarcity of radio sources with strong shocks and the need for distributed heating[81]. In the transient active phase, jets inflate cavities and shocks are the main heating mechanism. If the jets can cross the cluster core before this phase ends, then shock heating raises the entropy by a nearly uniform amount throughout the core, in agreement with the observed ~10 keV $cm^2$ entropy pedestal (Fig. 3). After the jets switch off, the radio lobes keep doing mechanical work on the intracluster medium by rising buoyantly. The thermalization of the hydrodynamic motions generated during the active phase and of the waves generated by the rising bubbles generates heat at various radii, even when the black hole is not actually accreting[44,45,49,77].

Heating is not the only mechanism by which AGN can prevent cold gas from accumulating at the centres of clusters. Jets and rising bubbles also lift low-entropy gas from the central region and transport it outside. Eventually this gas falls down again, but new bubbles are created and lift it up again[82]. This would explain the filaments of cold molecular gas detected around the cavities of the Perseus cluster[83].

## Black holes in galaxy evolution

Giant ellipticals have the same cooling flow problem as galaxy clusters, with even stronger limits on the amount of gas that can cool and form stars. The stellar populations, chemical abundances, and structural properties (that is, the absence of dense central light cusps[57]) of giant ellipticals indicate that little gas has fallen to the centre and made stars since these galaxies were formed.

The cooling flow problem is more severe within galaxies than within clusters because, even neglecting the hot gas in the halo, the final stages of the lives of massive stars return ~30–40% of the total stellar mass to the interstellar medium over the lifetime of the Universe[50,51,84]. Even a small fraction of the gas from dying massive stars would, if accreted, result in black holes much larger than the observational mass estimates.

The problem with applying to galaxies the same explanation as applied to clusters is that jets are usually collimated on galactic scales[85] (see, for example, M87). Therefore, they drill through the nearby gas and dump most of the energy outside the galaxies in which they are produced: the entropy shelf surrounding NGC 6166 (Fig. 3) indicates that heating has only been important at radii $r \geq 2$ kpc. Even in the Perseus cluster, the jets seem to not be inhibiting star formation in the central galaxy, which belongs to the 25% of cD galaxies that are blue[86]. The situation is even worse in galaxies that are not at the centres of clusters, because confining the jets is even more difficult for the less pressurized atmospheres of these galaxies. Without a confining working surface, jets dissipate their energy uselessly in intergalactic space[57] (for example, Cygnus A). However, there are counter-examples where the jets have caused turmoil in the hot gas on galactic scales—for example, Centaurus A, M84 and NGC 3801. Moreover, a jet may escape from its host galaxy and still transfer some of its energy to the interstellar medium. For example, the knots in the jet of M87 could be interpreted as evidence for interaction with the interstellar medium. Despite this problem, it is intriguing that the fraction of ellipticals hosting a radio source scales with $M_{BH}$ in the same way as does the estimated gas cooling rate[87], and that the time-averaged jet power matches the gas X-ray luminosity over two orders of magnitude in galaxy mass[88]. If jets fail to couple to the ambient gas and to keep it hot, cooling will eventually activate an optical AGN, which could heat the gas radiatively[51].

Gravitational heating, due to the mechanical work done by infalling clumps when they fall deep into the galaxies[89–91], also contributes to heating the gas, as do type I supernovae in small ellipticals[50,51,84]. However, these energy sources are unresponsive to changes in the radiative loss rate. Either they will heat the gas at a rate lower than the cooling rate, in which case the gas will eventually cool, or they will heat the gas at a rate higher than the cooling rate, in which case they will drive an outflow. The gas may be in an outflow in lower-mass ellipticals (the 'cuspy' ellipticals of Fig. 1), where discrete sources dominate the X-ray emission and where, for this reason, we cannot generally detect any X-ray-emitting gas, but in giant ellipticals X-ray observations show that the gas is in hydrostatic equilibrium[57]. Gravitational heating and type I supernovae could, nevertheless, alleviate the burden on the AGN, which would only provide the difference between the heating rate needed to keep the gas in equilibrium and the heating rate provided by these other sources.

## Remaining issues

The strongest evidence for black hole feedback is in galaxy clusters, but we still lack a sufficient understanding of the processes that transfer energy from AGN to the surrounding gas and thermalize the hydrodynamic disturbances excited by expanding jets and raising bubbles. Standard viscosity, turbulent viscosity, the stretching and tearing of magnetic field lines, and cosmic rays could all contribute to heat and/or lift the intracluster medium. The statement that black holes self-regulate to the accretion rate that is required to offset cooling is a valid first approximation, but some gas does cool and flow onto the central galaxies of clusters[83], although at a very low rate compared to predictions for pure cooling flow models. In 25% of all clusters, this gas reactivates star formation, leading to blue-core cD galaxies[86]. A major challenge for theoretical models and computer simulations is to understand in quantitative detail why real clusters depart from an 'ideal' feedback loop that is 100% efficient in suppressing cooling and star formation.

The interaction of radio galaxies with their own interstellar media is much less clear than the interaction with the intracluster medium. In the case of radiative feedback, the basic physics of the interaction with the interstellar medium are much better understood. The main open problem is rather whether radiative feedback can deliver the energy required for the 'maintenance' of individual ellipticals without exceeding the observational limits on the fraction containing an AGN.

The greatest uncertainty is the role of quasar winds in quenching star formation. This is because the masses of the winds detected spectroscopically are uncertain by more than one order of magnitude. Improving the current estimates for the masses, length scales, and temperature structure of the winds at all redshifts is the critical observational challenge. We also need to understand better the properties of galaxies in transition from the blue to the red population.

Finally, we note that it is computer simulations that indicate the need for quasar quenching, but these simulations are based on uncertain models for star formation and the physics of the interstellar medium. Progress in our understanding of these processes and higher resolution simulations will be necessary before we can conclude that quasar feedback is in fact needed, particularly in lower-mass ellipticals where the decline of the star formation rate occurs on a longer timescale.

1. Kormendy, J. in *The Nearest Active Galaxies* (eds Beckman, J., Colina, L. & Netzer, H.) 197–218 (Consejo Superior de Investigaciones Científicas, Madrid, 1993).

2. Magorrian, J. *et al.* The demography of massive dark objects in galaxy centers. *Astron. J.* **15,** 2285–2305 (1998).

3. Marconi, A. & Hunt, L. The relation between black hole mass, bulge mass, and near-infrared luminosity. *Astrophys. J.* **589,** 21–24 (2003).

4. Ferrarese, L. & Merritt, D. A fundamental relation between supermassive black holes and their host galaxies. *Astrophys. J.* **539,** 9–12 (2000).

5. Gebhardt, K. *et al.* A relationship between nuclear black hole mass and galaxy velocity dispersion. *Astrophys. J.* **539,** 13–16 (2000).

6. Cattaneo, A. & Bernardi, M. The quasar epoch and the stellar ages of early-type galaxies. *Mon. Not. R. Astron. Soc.* **344,** 45–52 (2003).

7. Hopkins, P. F. Determining the properties and evolution of red galaxies from the quasar luminosity function. *Astrophys. J.* **163** (Suppl.), 50–79 (2006).

8. Lynden-Bell, D. Galactic Nuclei as Collapsed Old Quasars. *Nature* **223,** 690–694 (1969).

Bardeen, J. M. Kerr metric black holes. *Nature* **226,** 64–65 (1970).

9. Krolik, J. H. *Active Galactic Nuclei: From the Central Black Hole to the Galactic Environment* (Princeton Univ. Press, 1999).

10. Cattaneo, A. & Best, P. N. On the jet contribution to the AGN cosmic energy budget. *Mon. Not. R. Astron. Soc.* **395,** 518–523 (2009).

11. White, S. D. M. & Rees, M. J. Core condensation in heavy halos — a two-stage theory for galaxy formation and clustering. *Mon. Not. R. Astron. Soc.* **183,** 341–358 (1978).

12. Blumenthal, G. R., Faber, S. M., Primack, J. R. & Rees, M. J. Formation of galaxies and large-scale structure with cold dark matter. *Nature* **311,** 517–525 (1984).

13. White, S. D. & Frenk, C. S. Galaxy formation through hierarchical clustering. *Astrophys. J.* **379,** 52–79 (1991).

14. Rees, M. J. & Ostriker, J. P. Cooling, dynamics and fragmentation of massive gas clouds — clues to the masses and radii of galaxies and clusters. *Mon. Not. R. Astron. Soc.* **179,** 541–559 (1977).

15. Silk, J. On the fragmentation of cosmic gas clouds. I — The formation of galaxies and the first generation of stars. *Astrophys. J.* **211,** 638–648 (1977).

16. Binney, J. The physics of dissipational galaxy formation. *Astrophys. J.* **215,** 483–491 (1977).

17. Keres, D. *et al.* How do galaxies get their gas? *Mon. Not. R. Astron. Soc.* **363,** 2–28 (2005).

18. Dekel, A. & Birnboim, Y. Galaxy bimodality due to cold flows and shock heating. *Mon. Not. R. Astron. Soc.* **368,** 39–55 (2006).

19. Fall, S. M. & Efstathiou, G. Formation and rotation of disc galaxies with haloes. *Mon. Not. R. Astron. Soc.* **193,** 189–206 (1980).

20. Steinmetz, M. Numerical simulations of galaxy formation. *Astrophys. Space Sci.* **269,** 513–532 (1999).

21. Dekel, A. *et al.* Cold streams in early massive hot haloes as the main mode of galaxy formation. *Nature* **457,** 451–454 (2009).

22. Toomre, A. & Toomre, J. Galactic bridges and tails. *Astrophys. J.* **178,** 623–666 (1972).

23. Sanders, D. B. *et al.* Ultraluminous infrared galaxies and the origin of quasars. *Astrophys. J.* **325,** 74–91 (1988).

24. Nardini, E. *et al.* Spectral decomposition of starbursts and AGNs in 5–8 micron Spitzer IRS spectra of local ULIRGs. *Mon. Not. R. Astron. Soc.* **385,** 130L–134L (2008).

25. Hutchings, J. B. & Campbell, B. Are QSOs activated by interactions between galaxies? *Nature* **303,** 584–588 (1983).

26. Dunlop, J. S. *et al.* Quasars, their host galaxies and their central black holes. *Mon. Not. R. Astron. Soc.* **340,** 1095–1135 (2003).

27. Bennert, N. *et al.* Evidence for merger remnants in early-type host galaxies of low-redshift QSOs. *Astrophys. J.* **677,** 846–867 (2008).

28. Arav, N. *et al.* HST STIS observations of PG0946+301: the highest quality spectrum of a BALQSO. *Astrophys. J.* **561,** 118–130 (2001).

29. de Kool, M. *et al.* Keck HIRES observations of the QSO FIRST J104459.6+365605: evidence for a large-scale outflow. *Astrophys. J.* **548,** 609–623 (2001).

30. Reeves, J. N., O'Brien, P. T. & Ward, M. J. A massive X-ray outflow from the quasar PDS456. *Astrophys. J.* **593,** 65–68 (2004).

31. Chartas, G., Brandt, W. N., Gallagher, S. C. & Proga, D. XMM-Newton and Chandra spectroscopy of the variable high-energy absorption of PG 1115+080: refined ouflow constraints. *Astron. J.* **133,** 1849–1860 (2007).

32. Nesvadba, N. P. H. *et al.* Evidence for powerful AGN winds at high redshift: dynamics of galactic outflows in radio galaxies during the "Quasar Era". *Astron. Astrophys.* **491,** 407–424 (2008).

33. Proga, D. in 267–276 (ASP Conf. Ser. Vol. 373, Astronomical Society of the Pacific, 2007).

34. Silk, J. Ultraluminous starbursts from supermassive black hole-induced outflows. *Mon. Not. R. Astron. Soc.* **364,** 1337–1342 (2005).

35. Silk, J. & Rees, M. J. Quasars and galaxy formation. *Astron. Astrophys.* **331,** 1L–4L (1998).

36. King, A. Black holes, galaxy formation, and the $M_{\rm BH}$-σ relation. *Astrophys. J.* **596,** 27–29 (2003).

37. Murray, N., Quataert, E. & Thompson, T. A. On the maximum luminosity of galaxies and their central black holes: feedback from momentum-driven winds. *Astrophys. J.* **618,** 569–585 (2005).

38. Fabian, A. C., Celotti, A. & Erlund, M. C. Radiative pressure feedback by a quasar in a galactic bulge. *Mon. Not. R. Astron. Soc.* **373,** 16L–20L (2006).

39. Robertson, B. *et al.* The evolution of the $M_{\rm BH}$-σ relation. *Astrophys. J.* **641,** 90–102 (2006).

40. Springel, V., Di Matteo, T. & Hernquist, L. Black holes in galaxy mergers: the formation of red elliptical galaxies. *Astrophys. J.* **620,** 79–82 (2005).

41. Hopkins, P. F. *et al.* A unified, merger-driven model of the origin of starbursts, quasars, the cosmic X-ray background, supermassive black holes, and galaxy spheroids. *Astrophys. J.* **163** (Suppl.), 1–49 (2006).

42. Thomas, D., Maraston, C., Bender, R. & Mendes de Oliveira, C. The Epochs of Early-Type Galaxy Formation as a Function of Environment. *Astrophys. J.* **621,** 673–694 (2005).

43. Böhringer, H. *et al.* A ROSAT HRI study of the interaction of the X-ray-emitting gas and radio lobes of NGC 1275. *Mon. Not. R. Astron. Soc.* **264,** 25L–28L (1993).

44. Forman, W. *et al.* Reflections of active galactic nucleus outbursts in the gaseous atmosphere of M87. *Astrophys. J.* **635,** 894–906 (2005).

45. Fabian, A. C. *et al.* A very deep Chandra observation of the Perseus cluster: shocks, ripples and conduction. *Mon. Not. R. Astron. Soc.* **366,** 417–428 (2006).

46. Rafferty, D. A., McNamara, B. R., Nulsen, P. E. J. & Wise, M. W. The feedback-regulated growth of black holes and bulges through gas accretion and starbursts in cluster central dominant galaxies. *Astrophys. J.* **652,** 216–231 (2006).

47. Churazov, E., Sunyaev, R., Forman, W. & Boehringer, H. Cooling flows as a calorimeter of active galactic nucleus mechanical power. *Mon. Not. R. Astron. Soc.* **332,** 729–734 (2002).

48. Cattaneo, A. & Teyssier, R. AGN self-regulation in cooling flow clusters. *Mon. Not. R. Astron. Soc.* **376,** 1547–1556 (2007).

49. Ruszkowski, M., Brueggen, M. & Begelman, M. Cluster heating by viscous dissipation of sound waves. *Astrophys. J.* **611,** 158–163 (2004).

50. Binney, J. & Tabor, G. Evolving cooling flows. *Mon. Not. R. Astron. Soc.* **276,** 663–678 (1995).

51. Ciotti, L. & Ostriker, J. P. Radiative feedback from massive black holes in elliptical galaxies: AGN flaring and central starburst fueled by recycled gas. *Astrophys. J.* **665,** 1038–1056 (2007).

52. Peterson, J. R. & Fabian, A. C. X-ray spectroscopy of cooling clusters. *Phys. Rep.* **427,** 1–39 (2006).

53. McNamara, B. R. & Nulsen, P. E. J. Heating hot atmospheres with active galactic nuclei. *Annu. Rev. Astron. Astrophys.* **45,** 117–175 (2007).

54. Croton, D. *et al.* The many lives of active galactic nuclei: cooling flows, black holes and the luminosities and colours of galaxies. *Mon. Not. R. Astron. Soc.* **365,** 11–28 (2006).

55. Bower, R. *et al.* Breaking the hierarchy of galaxy formation. *Mon. Not. R. Astron. Soc.* **370,** 645–655 (2006).

56. Cattaneo, A. *et al.* Modelling the galaxy bimodality: shutdown above a critical halo mass. *Mon. Not. R. Astron. Soc.* **370,** 1651–1665 (2006).

57. Kormendy, J., Fisher, D. B., Cornell, M. E. & Bender, R. Structure and formation of elliptical and spheroidal galaxies. *Astrophys. J. Suppl. Ser.* (in the press); preprint at ⟨http://arxiv.org/abs/0810.1681⟩ (2008).

58. Lutz, D. *et al.* Star formation in the hosts of high-z QSOs: evidence from Spitzer PAH detections. *Astrophys. J.* **684,** 853–861 (2008).

59. Heckman, T. M. & Kauffmann, G. The host galaxies of AGN in the Sloan Digital Sky Survey. *N. Astron. Rev.* **50,** 677–684 (2006).

60. Tremonti, C. A., Moustakas, J. & Diamond-Stanic, A. M. The discovery of 1000 km $s^{-1}$ outflows in massive poststarburst galaxies at z~0.6. *Astrophys. J.* **663,** 77–80 (2007).

61. Schawinski, K. *et al.* Observational evidence for AGN feedback in early-type galaxies. *Mon. Not. R. Astron. Soc.* **382,** 1415–1431 (2007).

62. Quintero, A. *et al.* Selection and photometric properties of K+A galaxies. *Astrophys. J.* **602,** 190–199 (2004).

63. Chen, Y. *et al.* Statistics of X-ray observables for the cooling-core and non-cooling core galaxy clusters. *Astron. Astrophys.* **466,** 805–812 (2007).

64. Ponman, T. J., Bourner, P. D. J., Ebeling, H. & Boehringer, H. A. ROSAT survey of Hickson's compact galaxy groups. *Mon. Not. R. Astron. Soc.* **283,** 690–708 (1996).

65. Edge, A. C. & Stewart, G. C. EXOSAT observations of clusters of galaxies. I — The X-ray data.II — X-ray to optical correlations. *Mon. Not. R. Astron. Soc.* **252,** 414–441 (1991)

66. Evrard, A. E. & Henry, J. P. Expectations for X-ray cluster observations by the ROSAT satellite. *Astrophys. J.* **383,** 95–103 (1991).

67. Kaiser, N. Evolution of clusters of galaxies. *Astrophys. J.* **383,** 104–111 (1991).

68. Lloyd-Davies, E. J., Ponman, T. J. & Cannon, D. B. The entropy and energy of intergalactic gas in galaxy clusters. *Mon. Not. R. Astron. Soc.* **315,** 689–702 (2000).

69. Valageas, P. & Silk, J. The entropy history of the universe. *Astron. Astrophys.* **350,** 725–742 (1999).

70. Oh, S. P. & Benson, A. J. Entropy injection as a global feedback mechanism. *Mon. Not. R. Astron. Soc.* **342,** 664–672 (2003).

71. McCarthy, I. G., Babul, A., Bower, R. G. & Balogh, M. L. Towards a holistic view of the heating and cooling of the intracluster medium. *Mon. Not. R. Astron. Soc.* **386,** 1309–1331 (2008).

72. Burns, J. O. The radio properties of cD galaxies in Abell clusters. I — an X-ray selected sample. *Astron. J.* **99,** 14–30 (1990).

73. Best, P. N. *et al.* On the prevalence of radio-loud active galactic nuclei in brightest cluster galaxies: implications for AGN heating of cooling flows. *Mon. Not. R. Astron. Soc.* **379,** 894–908 (2007).

74. Blandford, R. D. & Begelman, M. C. On the fate of gas accreting at a low rate on to a black hole. *Mon. Not. R. Astron. Soc.* **303,** 1L–5L (1999).

75. Dunn, R. J. H. & Fabian, A. C. Investigating AGN heating in a sample of nearby clusters. *Mon. Not. R. Astron. Soc.* **373,** 959–971 (2006).

76. Nusser, A., Silk, J. & Babul, A. Suppressing cluster cooling flows by self-regulated heating from a spatially distributed population of active galactic nuclei. *Mon. Not. R. Astron. Soc.* **373,** 739–746 (2006).

77. Binney, J., Bibi, F. A. & Omma, H. Bubbles as tracers of heat input to cooling flows. *Mon. Not. R. Astron. Soc.* **377,** 142–146 (2007).

78. Bondi, H. On spherically symmetrical accretion. *Mon. Not. R. Astron. Soc.* **112,** 195–204 (1952).

79. Allen, S. W. *et al.* The relation between accretion rate and jet power in X-ray luminous elliptical galaxies. *Mon. Not. R. Astron. Soc.* **372,** 21–30 (2006).

80. Magliocchetti, M. & Brueggen, M. The interplay between radio galaxies and cluster environment. *Mon. Not. R. Astron. Soc.* **379,** 260–274 (2007).

81. Voit, G. M. & Donahue, M. An observationally motivated framework for AGN heating of cluster cores. *Astrophys. J.* **634,** 955–963 (2005).

82. Mathews, W. G. & Brighenti, F. Creation of X-ray cavities in clusters with cosmic rays. *Astrophys. J.* **660,** 1137–1145 (2006).

83. Salomé, P. *et al.* Cold molecular gas in the Perseus cluster core. *Astron. Astrophys.* **454,** 437–445 (2006).

84. Mathews, W. G. & Baker, J. C. Galactic winds. *Astrophys. J.* **170,** 241–260 (1971).

85. Helmboldt, J. F., Taylor, G. B., Walker, R. C. & Blandford, R. D. A statistical description of AGN jet evolution from the VLBA Imaging and Polarimetry Survey (VIPS). *Astrophys. J.* **681,** 897–904 (2008).

86. Bildfell, C., Hoekstra, H., Babul, A. & Mahdavi, A. Resurrecting the red from the dead: optical properties of BCGs in X-ray luminous clusters. *Mon. Not. R. Astron. Soc.* **389,** 1637–1654 (2008).

87. Best, P. N. *et al.* The host galaxies of radio-loud active galactic nuclei: mass dependences, gas cooling and active galactic nuclei feedback. *Mon. Not. R. Astron. Soc.* **362,** 25-40(2005).

88. Best, P. N., Kaiser, C. M., Heckman, T. M. & Kauffmann, G. AGN-controlled cooling in elliptical galaxies. *Mon. Not. R. Astron. Soc.* **368,** 67L–71L (2007).

89. Dekel, A. & Birnboim, Y. Gravitational quenching in massive galaxies and clusters by clumpy accretion. *Mon. Not. R. Astron. Soc.* **383,** 119–138 (2008).

90. Khochfar, S. & Ostriker, J. P. O. Adding environmental gas physics to the semianalytic method for galaxy formation: gravitational heating. *Astrophys. J.* **680,** 54–69 (2008).

91. Naab, T., Johansson, P. H., Ostriker, J. P. & Efstathiou, G. Formation of early-type galaxies from cosmological initial conditions. *Astrophys. J.* **658,** 710–720 (2007).

92. Baldry, I. K. *et al.* Quantifying the bimodal color magnitude distribution of galaxies. *Astrophys. J.* **600,** 681–694 (2004).

93. Faber, S. M. *et al.* The centers of early-type galaxies with HST. IV. Central parameter relations. *Astron. J.* **114,** 1771–1796 (1997).

94. Naab, T., Khochfar, S. & Burkert, A. Properties of early type, dry galaxy mergers and the origin of massive elliptical galaxies. *Astrophys. J.* **636,** 81–84 (2006).

95. Cox, T. J. *et al.* The kinematic structure of merger remnants. *Astrophys. J.* **650,** 791–811 (2006).

96. Faber, S. M. *et al.* Galaxy luminosity functions to z~1 from DEEP2 and COMBO-17: implications for red galaxy formation. *Astrophys. J.* **665,** 265–294 (2007).

97. Khalatyan, A. *et al.* Is AGN feedback necessary to form red elliptical galaxies? *Mon. Not. R. Astron. Soc.* **387,** 13–30 (2008).

98. Donahue, M., Horner, D. J., Cavagnolo, K. W. & Voit, G. M. Entropy profiles in the cores of cooling flow clusters of galaxies. *Astrophys. J.* **643,** 730–750 (2006).

**Acknowledgements** A.C. thanks his wife A. Fylaktou for assistance in making the Review readable for a broader audience.

**Author Contributions** A.C. initiated the project, wrote the first draft, and had editorial control throughout. He chose to have many co-authors to show that the Review reflects consensus within the field. S.M.F. made a major contribution to the structure and content of the Review. J.B., A.D., J.K., and R.M. participated extensively in the writing of the manuscript, A.B., P.B., and M.B. contributed significantly to individual sections. A.K., A.M., and P.B. produced Figures 2, 3, and 4, respectively. The other authors contributed mainly by providing comments on drafts and by participating in scientific discussions in connection with: X-ray observations of galaxy clusters (A.C.F.), galaxy formation (C.S.F., M.S.), the interaction of radiation with the interstellar medium (H.N.), the interaction of quasar winds with the interstellar medium (J.S.), quasar winds and host galaxies (L.W.).

**Author Information** Reprints and permissions information is available at www.nature.com/reprints. Correspondence should be addressed to A.C. (acattaneo@aip.de).

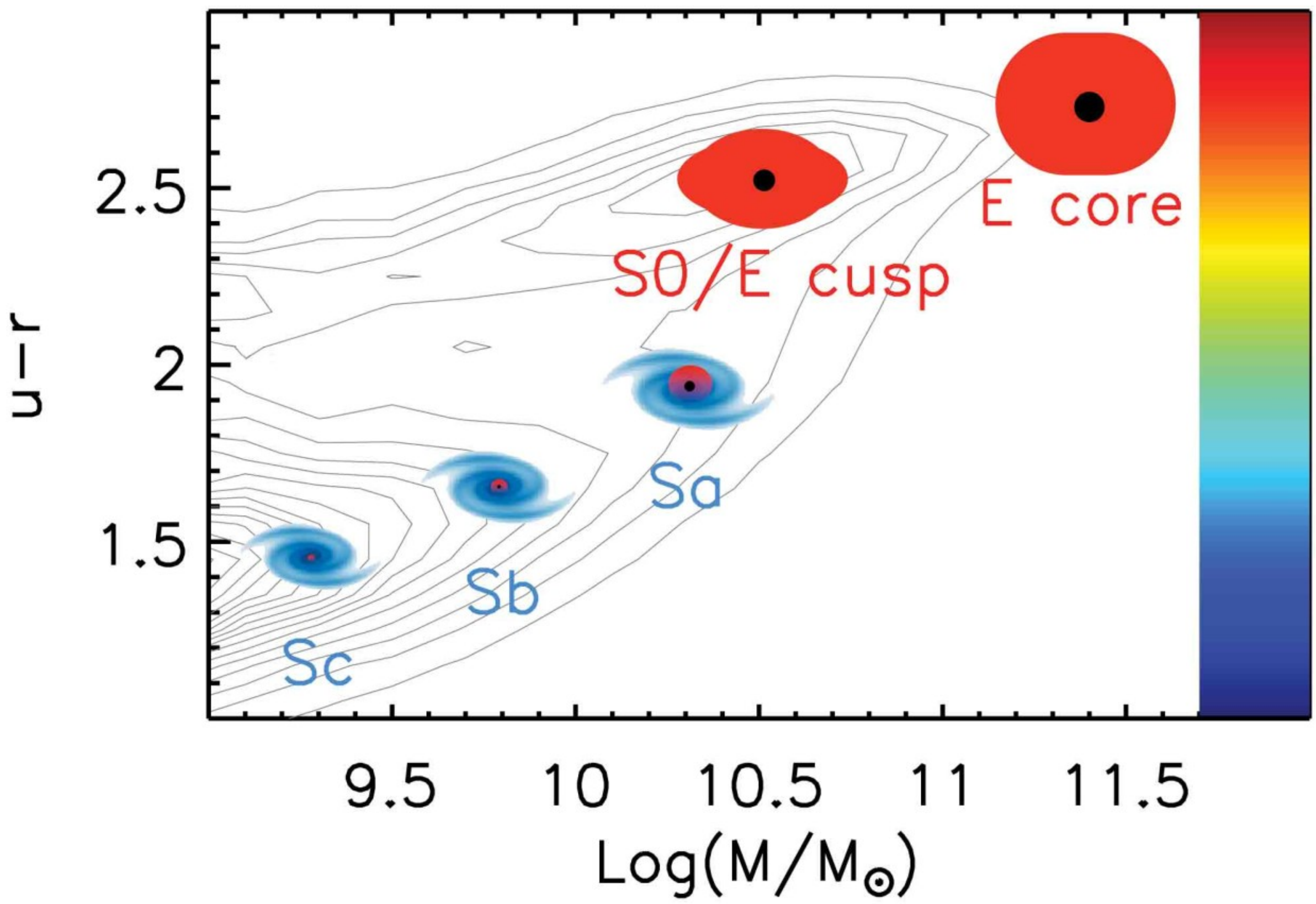

**Figure 1 The galaxy bimodality.** The contours show the galaxy distribution on a stellar mass ($M_{gal}$) - colour diagram (ref 92). The difference between ultraviolet luminosity and red luminosity, quantified by the magnitude difference $u - r$, is a colour indicator; larger values of $u - r$ correspond to redder galaxies. The colour bar has been inserted to convey this notion visually and has no quantitative meaning. Galaxies are classified into two main types: spirals that mainly grew through gas accretion ('S', shown in blue) and ellipticals that mainly grew through mergers with other galaxies ('E', shown in red). 'S0' galaxies are an intermediate type, but we assimilate them to ellipticals. Spirals have central bulges, shown in red, that resemble miniature ellipticals. All ellipticals and bulges within spirals contain a central black hole, shown with a black dot. Moreover, ellipticals and bulges within spirals have the same black-hole mass to stellar mass ratio, of the order of 0.1%. This is why we call them 'bulges' indiscriminately. In contrast, there is no connection between masses of black holes and masses of disks (the galactic component shown in blue). Spirals and ellipticals are separated by a colour watershed at $u - r \approx 2$ and a mass watershed at $M_{gal} \approx M^* \approx 10^{10.5} M_\odot$ (ref. 92). $M^*$ is of the order of $f_b M_{crit}$, where $M_{crit} \approx 10^{12} M_\odot$ is the critical halo mass for gas accretion and $f_b \approx 0.17$ is the cosmic baryon fraction. Spirals form a sequence where the bulge-to-disk ratio tends to grow with $M_{gal}$ (Sc, Sb, Sa). Ellipticals have two subtypes[57,93]: giant ellipticals with smooth low-density central cores formed in mergers of galaxies that have long finished their gas ('E core')[94] and lower-mass ellipticals with steep central light cusps formed in mergers of galaxies that still have gas ('E cusp')[95]. Whereas core ellipticals formed all their stars over a short time span at high redshift[42], the formation of the lower-mass cuspy ellipticals from the 'quenching' and reddening of blue galaxies continues to low redshift[96].

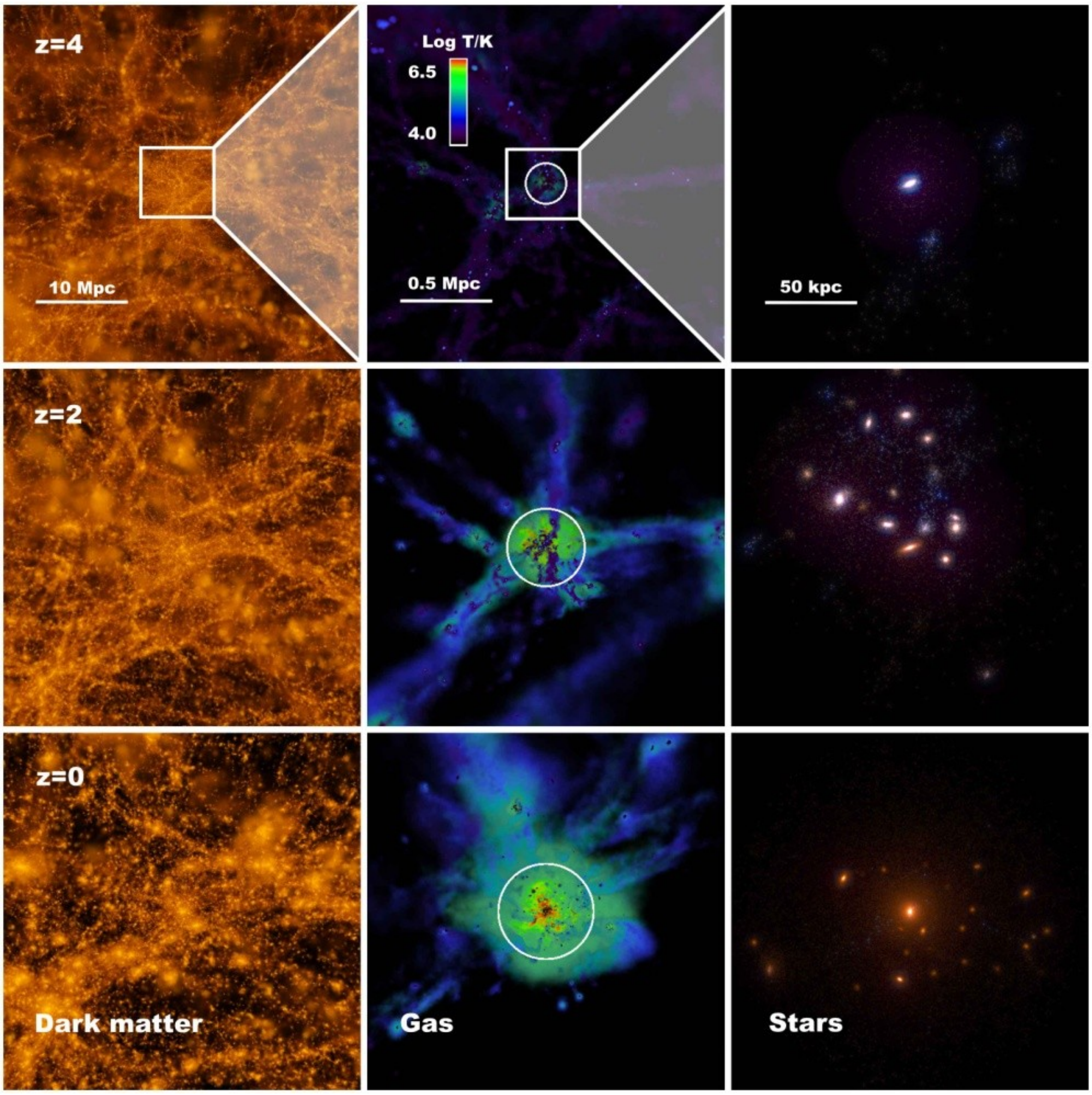

**Figure 2 A computer simulation of the formation of an elliptical galaxy**. The nine panels illustrate the formation of an elliptical galaxy[97] by showing how the dark matter (left column), the gas (centre column) and the stars (right column) are distributed at three epochs in the expansion of the Universe: when the Universe was 1/5 of its current size (redshift $z = 4$), when the Universe was 1/3 of its current size ($z = 2$), and today ($z = 0$). The gravity of the dark matter dominates the evolution on large scales (left column). As time passes, the Universe becomes lumpier because the dark matter clumps via gravity into haloes (bright orange spots in the left panels). The centre column zooms into the region around and inside a halo to show what happens to the gas. The halo radius is shown as a white circle, and the gas is colour-coded according to its temperature: blue is cold, green (and red) is hot. Initially the halo is small, and the gas streams into the halo down to its centre in cold flows. When the halo reaches the critical mass $M_{crit} \approx 10^{12} M_{\odot}$ ($z = 2$), the gas begins to form a hot atmosphere (green); eventually, all the gas within the halo is hot ($z = 0$). The right column zooms in even further to show the visible galaxy formed by the gas fallen to the centre. The galaxy is initially a blue spiral ($z = 4$). It starts to become red when the halo gas starts to be hot ($z = 2$). By then, its halo has merged with neighbouring haloes to form a galaxy group. Mergers with companions eventually transform the galaxy into an elliptical ($z = 0$).

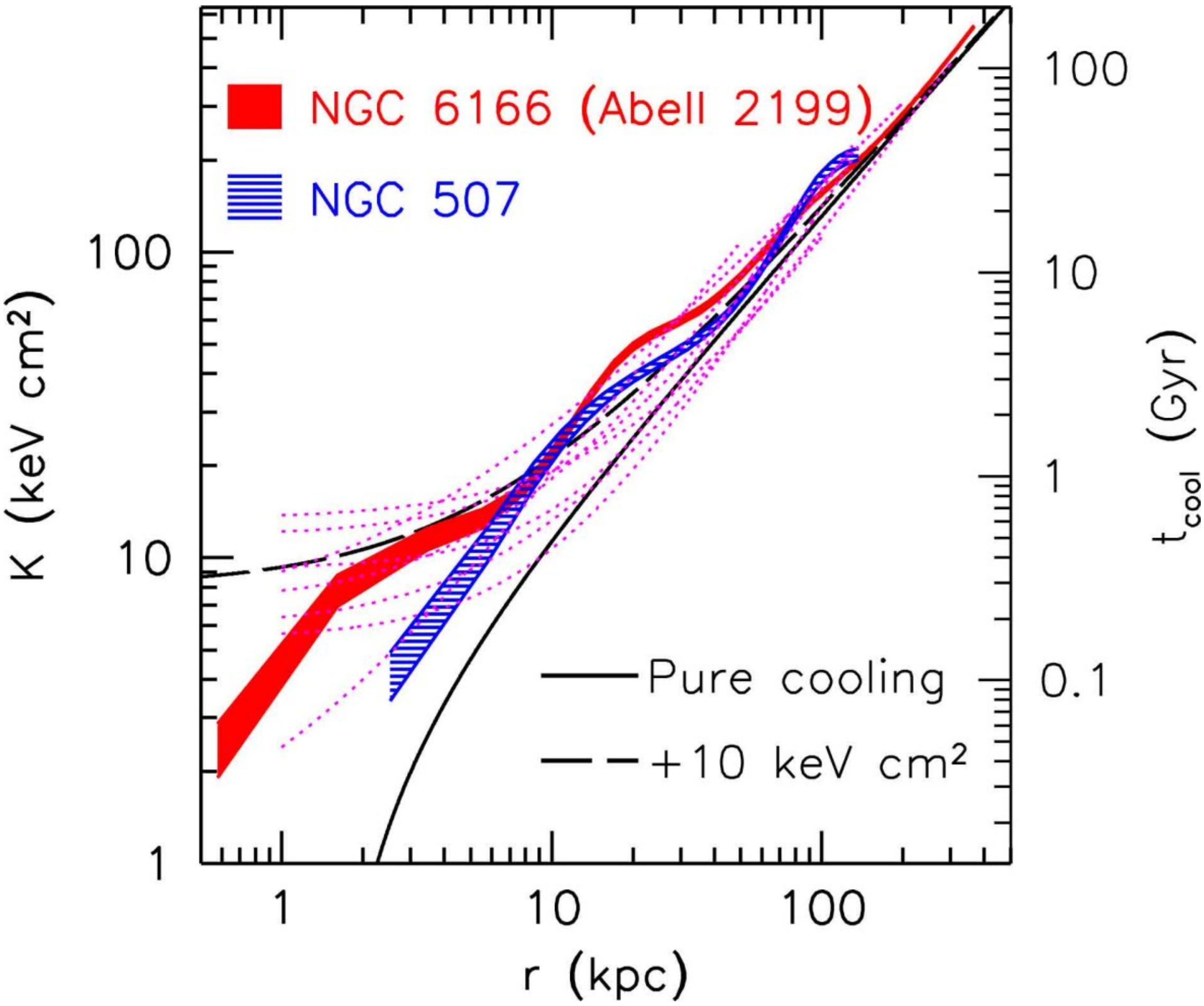

**Figure 3 The entropy of the intracluster medium in spherical shells of radius *r*.** We define entropy as $K = kTn^{1/(\gamma-1)}$, where $\gamma$ is the gas adiabatic index. For an ideal gas, this definition is related to the standard one, $\Delta S = \Delta Q/T$, by the transformation $s = \ln K^{1/(\gamma-1)}$ + const., where $s$ is the entropy $S$ per unit mass ($\Delta S$ is the entropy variation that corresponds to a heat injection $\Delta Q$). Observed entropy profiles of cool-core clusters[98] (red dotted lines) differ substantially from theoretical predictions for a pure cooling flow model (black solid line) but become broadly consistent with theoretical predictions if an entropy pedestal of 10 keV cm$^2$ is added to the latter (dashed black line). The discrepancy with predictions of pure cooling flow models is even larger in non-cool-core clusters, which have central entropies up to 700 keV cm$^2$ (for example, 3C 129). A new study by three of us (A.M., A.B. and A.C., unpublished results) resolves entropy profiles of NGC 507 and NGC 6166, the cD galaxy of A 2199, at small radii and finds that the pedestal is actually a shelf. The entropy decreases again at small radii. Both galaxies have half-light radii of ~10 kpc. We also show the radiative cooling times $t_{\rm cool}$ that approximately correspond to the entropies on the $y$-axis of the diagram. The dependence of $t_{\rm cool}$ on $K$, $t_{\rm cool} = 3/2K^{3/2}/(\Lambda\sqrt{kT})$ for $\gamma = 5/3$, is stronger than that on $T$ because of the exponent 3/2 and because the range of entropies within a cluster and among clusters is much larger than the corresponding range of temperatures.

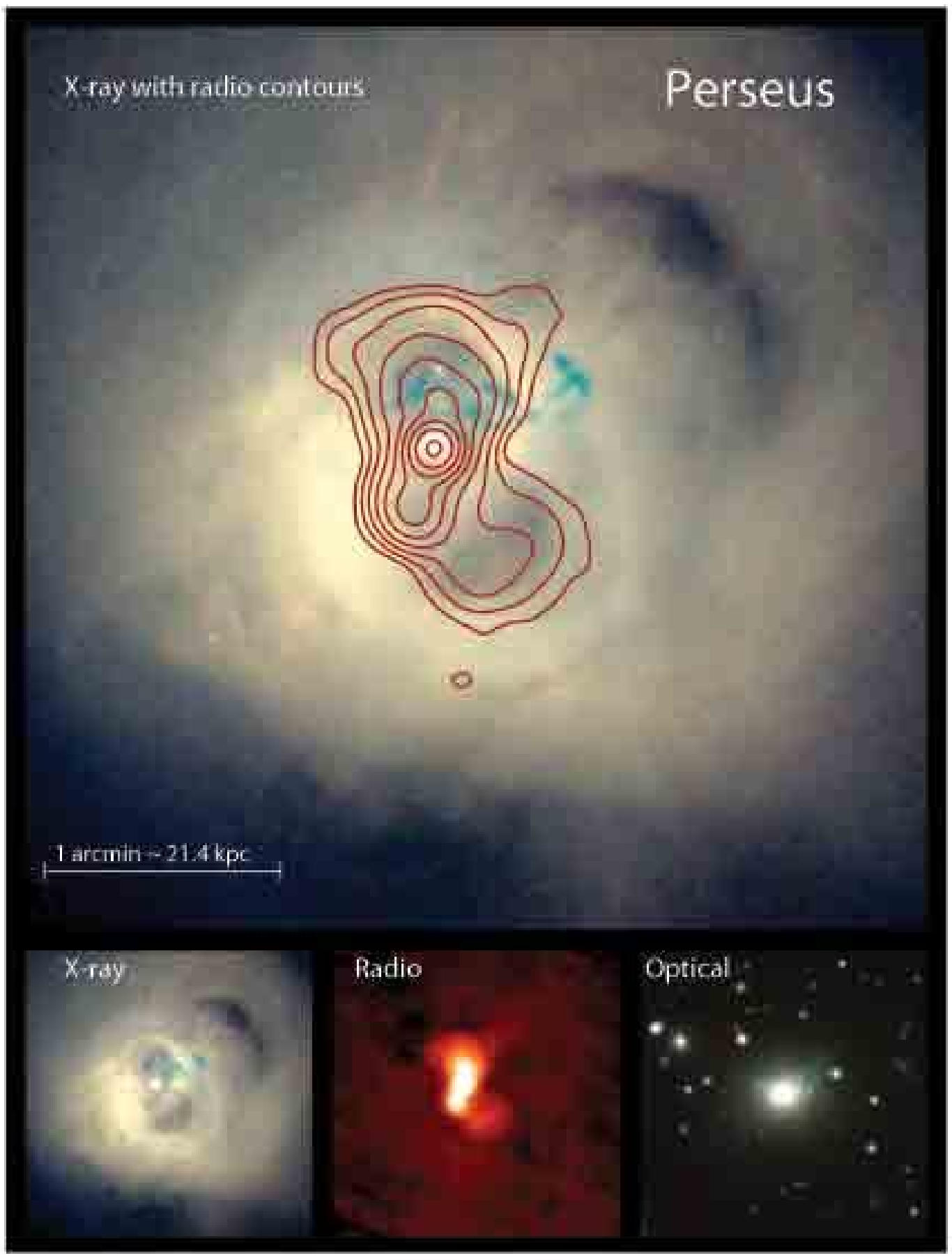

**Figure 4 Optical, radio and X-ray images of the Perseus cluster**. The optical image (bottom right panel) shows the Perseus galaxy cluster and its cD galaxy, the radio galaxy Perseus A. The radio images (bottom middle panel, and contours on main panel) show the lobes of relativistic synchrotron emitting plasma inflated by the jets that come out of the central nucleus. The X-ray images (main panel and bottom left panel) show the ambient hot gas (intracluster medium), colour-scaled such that higher-energy X-rays (hotter gas) are bluer and lower-energy X-rays (cooler gas) are redder (the X-ray and radio data are from ref 45). This is one of the most dramatic images of the cavities and ripples created by AGN in the surrounding gas. The regions of higher radio luminosity are dimmer in X-rays because the expanding lobes have displaced the ambient gas. The rims of the radio lobes are X-ray bright and are cooler than the surrounding gas. This is more obvious if one looks to the left of the spot where the jet directed downward terminates. A third X-ray cavity has the shape of an arch and is visible at the top right corner. This is called a 'ghost' cavity because it is invisible in radio, and it was generated during a previous burst of activity. The pink brush strokes running perpendicular to the arch are cold gas flowing around the cavity. The blue structure to the north of the active nucleus is due to absorption in an infalling system in front of Perseus A. This feature appears because higher-energy X-rays are more penetrating.

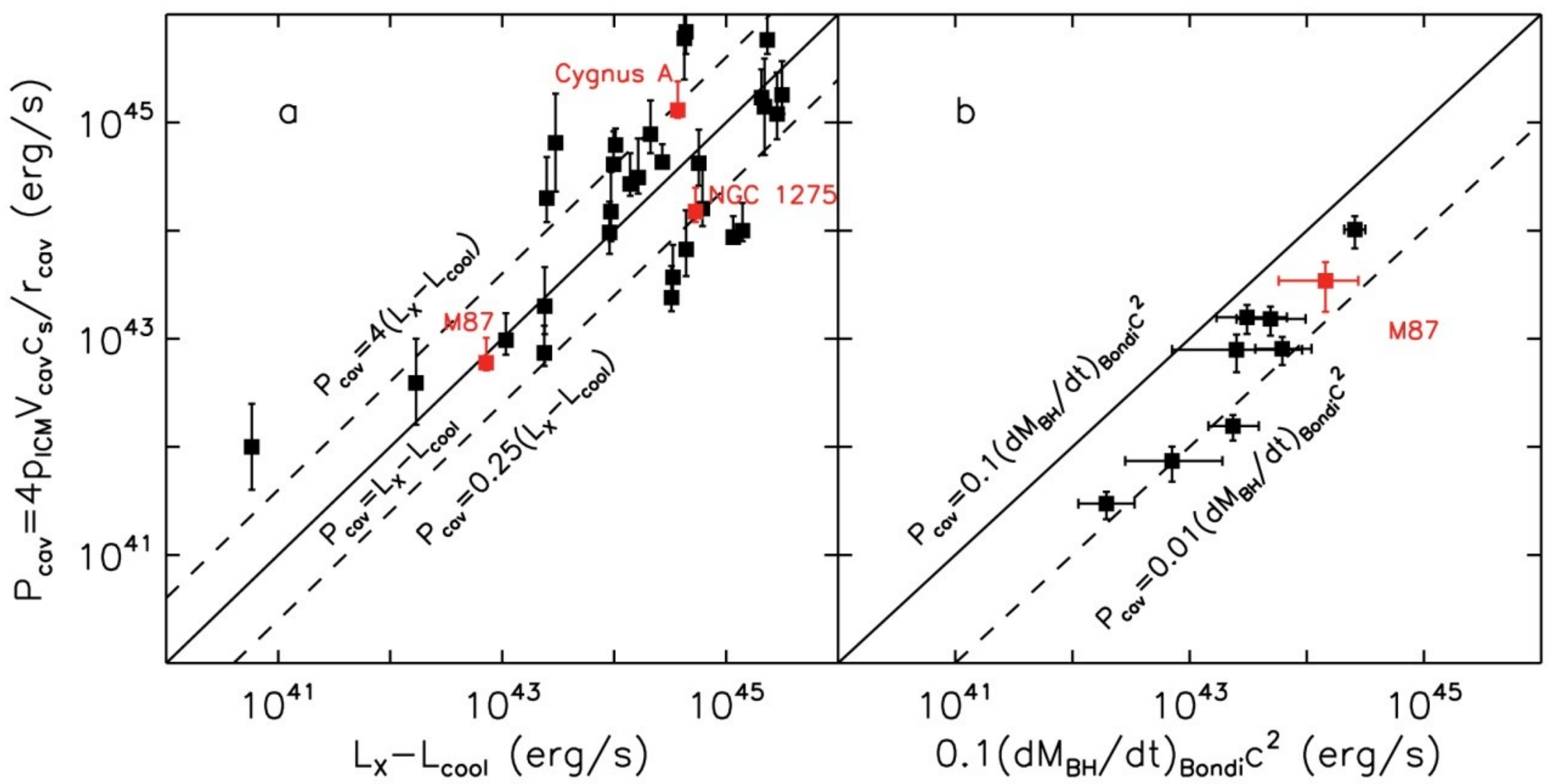

**Figure 5 Cooling, heating, and black hole accretion rates.** **a**, $P_{cav}$ measures the mechanical energy $E_{cav}$ used to create a cavity divided by the sound crossing time $r_{cav}/c_s$ (ref. 46). $L_X$ is the total X-ray luminosity of the intracluster medium inside the cooling radius (defined[46] as the radius within which $t_{cool} < 7.7$ Gyr). $L_{cool}$ is the X-ray luminosity of the gas that is actually cooling, computed by modelling X-ray and ultraviolet spectra. In fact, $L_{cool}$ is negligible compared to $L_X$ because the gas-cooling rate is low. If $E_{cav} = 4p_{ICM}V_{cav}$, then $P_{cav} = L_X$ to within a factor of four. This means that the jet energy is about equal to the energy that is needed to offset cooling. **b**, $P_{cav}$ ranges[79] from 10% to 100% of the accretion power determined from the Bondi spherical accretion model[78]. This shows that the Bondi model is a reasonable description of the accretion of hot gas by a supermassive black hole. The different ordinates of M87 in **a** and **b** show that the published error bars underestimate the real uncertainty on $P_{cav}$. Error bars, $1\sigma$.